\documentclass[10pt,conference]{IEEEtran}
\usepackage[utf8]{inputenc}

\usepackage{tabularx}
\usepackage{booktabs}
\usepackage{soul}
\usepackage{xcolor}
\usepackage{xspace}
\usepackage{url}
\usepackage{graphicx}
\usepackage[caption=false]{subfig}  
\usepackage[export]{adjustbox}
\usepackage{amsmath}
\usepackage{amssymb}
\usepackage{textcomp}
\usepackage{listings}
\usepackage{cleveref}
\usepackage{algorithm}
\usepackage[noend]{algpseudocode}

\definecolor{codegreen}{rgb}{0,0.6,0}
\definecolor{codegray}{rgb}{0.5,0.5,0.5}
\definecolor{codepurple}{rgb}{0.58,0,0.82}
\definecolor{backcolour}{rgb}{0.95,0.95,0.92}

\lstdefinestyle{mystyle}{
    stringstyle=\color{deepgreen},
    backgroundcolor=\color{backcolour},   
    commentstyle=\color{codegreen},
    keywordstyle=\color{magenta},
    numberstyle=\tiny\color{codegray},
    stringstyle=\color{codepurple},
    basicstyle=\ttfamily\footnotesize,
    breakatwhitespace=false,         
    breaklines=true,                 
    captionpos=b,                    
    keepspaces=true,                 
    numbers=left,                    
    numbersep=5pt,                  
    showspaces=false,                
    showstringspaces=false,
    showtabs=false,                  
    tabsize=2
}

\newcommand{\ie}{\emph{i.e.,}\xspace}

\newcommand{\toolname}{\textsc{ReSplit}\xspace}

\title{\toolname: Improving the Structure of Jupyter Notebooks by Re-Splitting Their Cells}

\author{
\IEEEauthorblockN{Sergey Titov}
\IEEEauthorblockA{
    \textit{JetBrains Research}\\
    Saint Petersburg, Russia \\
    sergey.titov@jetbrains.com
}
\and
\IEEEauthorblockN{Yaroslav Golubev}
\IEEEauthorblockA{
    \textit{JetBrains Research}\\
    Saint Petersburg, Russia \\
    yaroslav.golubev@jetbrains.com
}
\and
\IEEEauthorblockN{Timofey Bryksin}
\IEEEauthorblockA{
    \textit{JetBrains Research}\\
    \textit{HSE University}\\
    Saint Petersburg, Russia \\
    timofey.bryksin@jetbrains.com
}
}

\begin{document}

\maketitle

\begin{abstract}
Jupyter notebooks represent a unique format for programming --- a combination of code and Markdown with rich formatting, separated into individual cells. We propose to perceive a Jupyter Notebook cell as a simplified and raw version of a programming function. Similar to functions, Jupyter cells should strive to contain singular, self-contained actions. At the same time, research shows that real-world notebooks fail to do so and suffer from the lack of proper structure.

To combat this, we propose \toolname, an algorithm for an automatic re-splitting of cells in Jupyter notebooks. The algorithm analyzes definition-usage chains in the notebook and consists of two parts --- merging and splitting the cells. We ran the algorithm on a large corpus of notebooks to evaluate its performance and its overall effect on notebooks, and evaluated it by human experts: we showed them several notebooks in their original and the re-split form. In 29.5\% of cases, the re-split notebook was selected as the preferred way of perceiving the code. We analyze what influenced this decision and describe several individual cases in detail.    
\end{abstract}

\section{Introduction}\label{sec:introduction}

In the 1980-s, Knuth described the idea of literate programming as integrating the code and the natural language~\cite{knuth1984literate}. Modern computational notebooks enable this idea by consisting of intertwined cells with either code or rich formatted text, which allows the resulting interactive documents to have a narrative~\cite{rule2018exploration}. This aspect made computational notebooks popular in such areas as data science~\cite{perkel2018jupyter} and education~\cite{johnson2020jupyter}, where code is just as important as its commentary. However, there is another side to their features --- computational notebooks contain a lot of code duplicates~\cite{koenzen2020code}, lack proper environments~\cite{wang2021restoring}, and have a very low rate of reproducibility~\cite{pimentel2019large}. Overall, for a medium that strives to be more comprehensive than pure code, notebooks frequently fail to live up to this goal~\cite{head2019managing, chattopadhyay2020s}.  

Currently, there exist several approaches to improve the comprehensibility of notebooks~\cite{kang2021toonnote}. One of the main ideas in this field is to automate the annotation of code in a notebook --- from simply tagging notebook cells~\cite{zhang2020coral} up to fully generating the annotation text~\cite{wang2021themisto}. Another possible approach is to introduce the history of code changes in cells or rearrange cells in order to make the notebook reproducible~\cite{kery2018interactions}.

In this work, we focus on one of the basic parts of a notebook --- its cells. The presence of logically separated code snippets that can be easily executed is one of the most crucial features of notebooks. Our hypothesis is that an ideal cell could be viewed as an equivalent of a singular action (or even as a \textit{proto-function}), but because of the very non-restrictive environment and the nature of the software development process, these singular actions become entangled, which results in incomprehensible, non-singular cells.

In order to solve this problem, we propose an approach that is similar to the \textit{Extract} refactoring~\cite{tsantalis2011identification} for code improvement. We present \toolname, a two-step algorithm that suggests a new re-splitting of the existing cells based on a number of heuristics. On the first step, the algorithm suggests merging some of the cells together, and on the second step, it suggests extracting certain code fragments from the original cells into new ones. We expect the new splitting of cells to better represent the singularity of their actions, which would improve the structure and the coherence of a notebook.

To test \toolname, we carried out two evaluations. Firstly, we ran the approach on a large corpus of notebooks to evaluate its performance and its effect on the characteristics of the dataset. 
In the second evaluation, we compare the original and the re-split notebooks from the standpoint of their perception by human experts. We asked the participants to read a small sample of notebooks and then asked several questions about the content of cells in both versions. As a result, we discovered that users prefer the re-split version of a notebook in 29.5\% of cases. The current shortcomings are merging cells that generate different output and significant changes of the formatting structure.
Without these problems, the ratio of selecting the re-split notebook significantly rises, which indicates that \toolname has potential to automatically improve the structure of notebooks in the future.

The source code of \toolname and supplementary materials are available: \url{https://github.com/JetBrains-Research/ReSplit}.
\section{Background}\label{sec:background}

\subsection{Previous Work on Jupyter Notebooks}

Several approaches exist that aim to improve the comprehensibility of Jupyter notebooks in various ways.
Workflow-related approaches are targeted at introducing some structure or discipline to the process of writing a notebook. This can simply mean providing some advice on how to write a notebook~\cite{rule2019ten} or introducing fully-fledged toolkits for documenting the whole process of creating a notebook~\cite{pimentel2016yin, samuel2018provbook}. 
A more automated approach is to generate annotations for notebooks. 
Wang et al.~\cite{wang2021themisto} generated annotations for cells with a machine learning algorithm combined with simple additions from a fixed set of documented functions. The approach proved to be efficient --- the experts scored the automatically annotated notebooks as more readable and informative. 

However, the comprehensibility of notebooks also suffers from the innate messiness of the notebook \textit{structure} --- individual cells could be clear, but their order could be problematic. Several works focused on the structure of the notebooks. Most works are dedicated to implementing some kind of a version control system~\cite{brachmann2020your}. 
Also, Head et al.~\cite{head2019managing} proposed an algorithm that allows users to choose \textit{the resulting cells}, and then filters out all redundant code that is not related to the selected cells. While this approach works very well, we wanted to take one step further and try to restructure the notebook automatically, without the user's input.

\subsection{Notebook Re-Splitting}

In this work, we focus on re-splitting the notebook into cells. We believe that reorganizing code in the notebook should help with its compressibility. In the already discussed work of Head et al.~\cite{head2019managing}, during the user study, the authors showed that organizing code in cells is a useful feature for notebook cleaning. Also, Dong et al.~\cite{dongsplitting} demonstrated that one of the common refactorings for notebooks is splitting certain cells (or parts of the cells) into separate scripts. 

We hypothesize that Jupyter notebook cells could be perceived as \textit{proto-functions}. What we mean is that when the programmers create cells, they are using the same principles as when they create functions: the code in a function must do one dedicated thing, must be reusable, and should be independent from other code. However, because of the looser nature of the medium, these requirements are often applied less strictly. 

This leads us to the idea that we could improve code quality in the notebooks by re-splitting cells into a more \textit{functional} style: clearer actions, more reusable and independent code. Additionally, such an approach can work incrementally --- the users do not need to finish a notebook or finalize the results to start restructuring it. 

\section{Approach}\label{sec:algorithm}

\toolname consists of two steps --- merging and splitting the cells of the given notebook. Both of these steps operate using the definition-usage (DEF-USE) chain analysis~\cite{kennedy1978use}. This approach could be seen as a simplified version of data flow graph analysis --- we are analysing all connections between object definitions and their usages. We extract all the chains of such definitions and usages for notebooks, and perform cell transformations based on a set of heuristics. In this section, we explain these heuristics and transformations, you can find detailed and illustrative examples in the project's repository.

\subsection{Creating Code Dependency Representation}

\begin{figure}[h]
    \centering
    \includegraphics[width=2.7in]{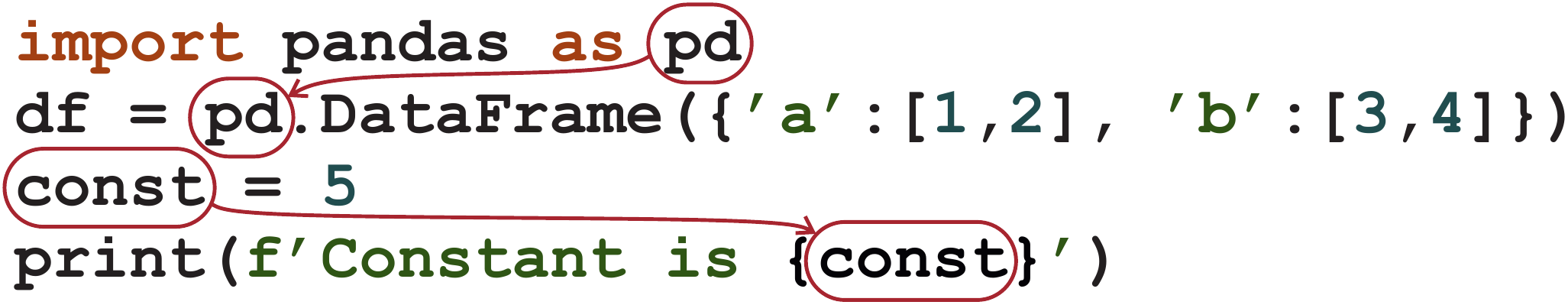}
    \caption{An example of a simple DEF-USE chain between several statements.}
    \label{fig:defuse}
    \vspace{-0.3cm}
\end{figure}

In order to obtain DEF-USE chains, we use the Python package called \textit{beniget},\footnote{Beniget: \url{https://github.com/serge-sans-paille/beniget}} which allows to extract DEF-USE chains for a given snippet. As a result, for each defined object, the package returns a sequence of nodes in the Abstract Syntax Tree (AST) where the object was used. 
This technique is very fast and allows us to process large corpora of notebooks. An example of a simple DEF-USE chain is presented in \Cref{fig:defuse}.

\subsection{Merging Cells}\label{sec:merge}

The main goal of merging the cells is to unite several statements into a single action. In terms of cells, we want to find consequent, related cells, and merge them into one. In order to do this, we introduce two assumptions about the cells, based on which we perform the merging. 

One thing we want to avoid during the merging process is turning big chunks of code into even bigger ones, so we introduce the requirement of the maximum size of the cell that we include in the merging. While there is no recommendation on what the optimal size of a cell should be, we decided to set it as 10 lines of code. This is slightly more than the suggested comfortable number of elements for human perception~\cite{miller1956magical} and less than the maximum recommendations in different programming style guides, like in the rule of 30~\cite{lippert2006refactoring}, where functions should take up less than 30 lines of code.

The second assumption has to do with different types of cells. In order to merge the cells, we build DEF-USE chains for the entire notebook. After we have extracted all the chains, we calculate the following two features for each cell:

\begin{itemize}
    \item \textbf{Intra-cell links}: the number of chain links between statements inside the cell (chains that are contained within the cell).
    \item \textbf{Inter-cell links}: the number of chain links to statements in other cells (chains that lead to other cells).
\end{itemize}

We hypothesise that cells in notebooks could be divided into two types: mostly intra-linked and mostly inter-linked. The intra-linked cells are usually operation-heavy. For example, such a cell can be a cell where the developer constructs features for a dataset and frequently uses temporary variables for intermediate results. At the same time, the inter-linked cells mostly operate with global objects that are used throughout the notebook --- like importing library functions or creating model objects that will later be used for prediction. 

Based on this assumption, we can use the ratio between intra-cell links and inter-cell links in the merging process. 
One problem that we want to solve with merging is the over-splitting of code: when developers use a separate cell for each individual statement. For example, this can happen when the user writes some complex piece of code line by line and uses separate cells for each line, or when they add imports one by one. We hypothesize that such cases could be defined in terms of intra- and inter-linked cells: consecutive cells of the same type are a part of one bigger action and we want to merge them. One way of doing it is to set the restriction on the permitted change of the ratio. In our algorithm, we calculate the \textit{inter-cell link ratio} as follows:

\[r_{inter}(c) = \frac{N_{inter\_cell\_links}}{N_{all\_links}}\]  

For each pair of cells that are candidates for merging, we compare the ratios before and after potential transformations and, in order to avoid merging different types of cells, we prohibit merges that cause a change in the cell ratio greater than a certain threshold. 

The resulting algorithm for merging the cells works as follows. We traverse the notebook, extract DEF-USE chains, and calculate $r_{inter}$ for each cell.
Then, for each pair of consecutive cells, we check for the following requirements: each cell is less than 5 lines and the potential change in the link ratio after merging is less than 0.1. If both these criteria are met, we merge these cells.

\subsection{Splitting Cells}

The goal of splitting cells is similar to that of the \textit{Extract Method} refactoring --- find discrete actions inside the cell and extract them into separate cells. To do so, we first need to detect these discrete actions. We do this with the following algorithm.
The first step is, once again, to extract all chains for each cell in the notebook. As a result, we can represent each cell as a set of Python statements linked by certain usage connections, \textit{e.g.}, if our code consists of three statements: 

\begin{lstlisting}[language=Python, xleftmargin=\parindent]
a = 2 + 2 
b = a / 2 
c = 16 * 2
\end{lstlisting}

\noindent then we can say that object \texttt{a} is defined in the first statement and is used in the second one, while the third statement is not connected to any of the first two. Having such a representation of cells, our algorithm seeks sequentially linked statements: if we are able to find a series of statements connected with DEF-USE chains with no other code between these statements, we mark the start and the end of the series as a potential place for splitting the cell.

The second step of the splitting part of the algorithm is building new cells. We start from the bottom of the original cell and collect statements into a new cell until we reach the potential place of splitting. If the number of already collected statements has exceeded the predefined threshold of a minimum of 3 lines to split, we perform the transformation, and if the number of statements is smaller, then we continue collecting the statements until the next potential split.
We chose to start from the bottom because we noticed that a considerable amount of cells have output statements at the end that could not be directly chained with previous statements. By starting with them, we are guaranteed to incorporate them in a bigger cell and not leave them to hang.

\section{Evaluation}\label{sec:evaluation}

To test \toolname, we perform two different evaluations. Firstly, we run the algorithm on a large corpus of code to test its performance and demonstrate how the algorithm affects the amount and length of cells in a large sample of notebooks. Secondly, we perform a human study to evaluate whether re-splitting the cells made the notebooks more comprehensible. 

\subsection{Dataset}

For our purposes, we took a dataset of 1,159,166 Jupyter notebooks from the work of Pimentel et al.~\cite{pimentel2019large}. Due to the complexity of our algorithm, we decided to reduce the size of the data. To do so, we decided to only keep the notebooks related to Data Science, since they allow us to retain the diversity of the data (prototypes, homeworks, reports, etc.).
We filtered the data by leaving only notebooks that contain at least one use of one of the following Python libraries --- \textit{sklearn}, \textit{pytorch}, \textit{tensorflow}, \textit{spacy}, and \textit{nltk}. We chose these libraries because they either provide crucial functions for most of the Data Science tasks or a full pipeline for some of them. This left us with 290,381 notebooks.

Additionally, based on the work of Koenzen et al.~\cite{koenzen2020code}, we decided to filter out code clones in the dataset. 
To detect clones, we used SourcererCC~\cite{sajnani2016sourcerercc}. We tokenized all the code cells in each notebook, and searched for clones using the similarity threshold of 80\%. As a result, we found and excluded 55,843 notebook duplicates, and the final dataset contained 234,538 notebooks. 

\subsection{Algorithm's Validation}

\begin{figure}[h]
    \centering
    \includegraphics[width=\columnwidth]{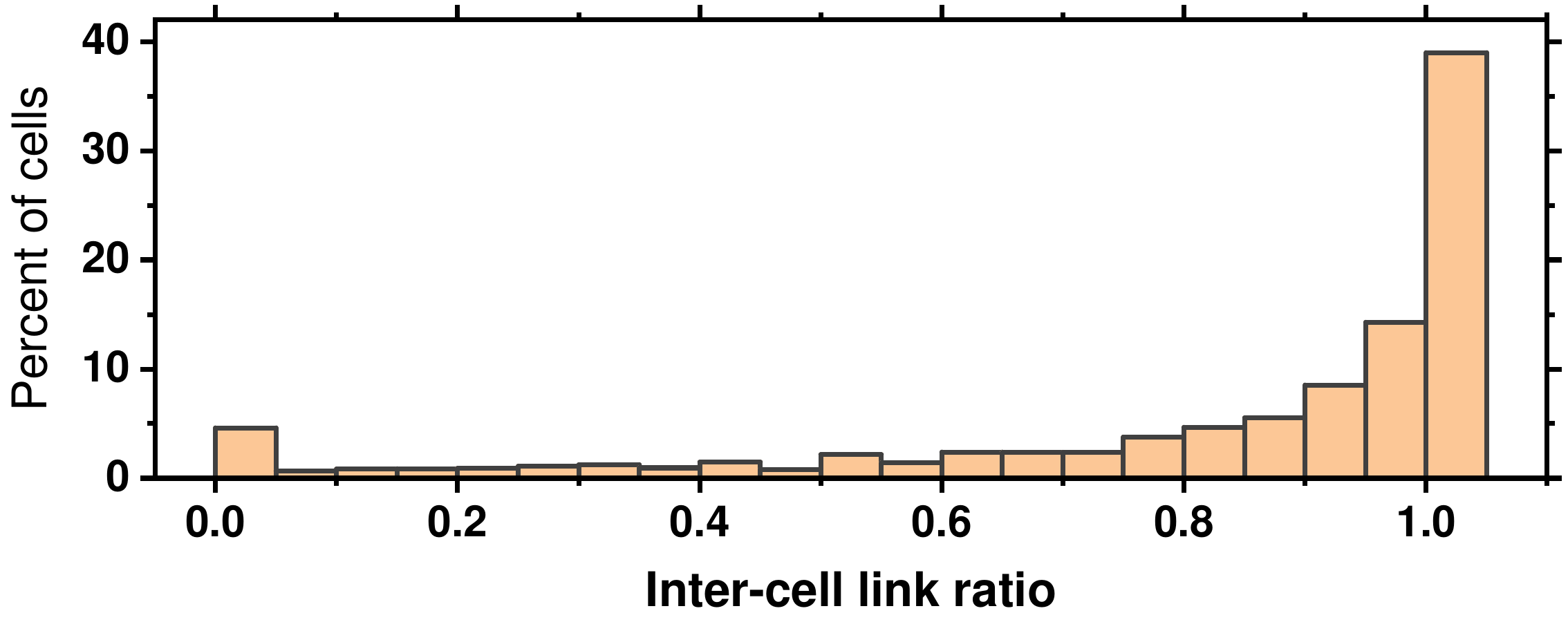}
    \caption{The distribution of inter-cell link ratio for all the cells in our dataset. }
    \label{fig:inter_dist}
\end{figure}

We start the analysis of the suggested algorithm by verifying one of our assumptions about the data that we made in \Cref{sec:merge}: we suggested that there are two types of cells --- mostly inter-linked and mostly intra-linked. Figure~\ref{fig:inter_dist} shows the inter-cell link ratio for all the cells in our dataset eligible for merging. We can see two noticeable spikes near 1 and 0 that demonstrate the hypothesized effect.

Now, let us take a look at the effect of applying \toolname on the characteristics of cells in the notebook. The mean number of cells in a notebook and the mean length of a cell in the dataset are shown in Table~\ref{table:mean_change}. The algorithm worked as expected --- merging caused an increase in the mean cell length and a decrease in the number of cells, while splitting caused the opposite effect. Both transformations applied together did not drastically change both characteristics, which shows that the merging and splitting processes are balancing each other out, thus not causing any drastic overhauls of the notebooks that could be uncomfortable for the user.

\begin{table}[h]
\centering
\begin{tabular}{ccc}
\toprule
\textbf{Transformations} & \textbf{Mean cell length} & \textbf{Mean number of cells} \\ \midrule
None (original dataset)      & 8.76           & 36.67        \\
Only merge                   & 11.32          & 28.95        \\
Only split                   & 7.91           & 39.89        \\
Both merge and split         & 9.47           & 34.52        \\
\bottomrule
\end{tabular}
\vspace{0.3cm}
\caption{The influence of the approach \\ on the characteristics of the dataset.}
\label{table:mean_change}
\vspace{-0.5cm}
\end{table}

\subsection{Human Evaluation}

Another hypothesis of ours was that re-splitting cells would help with the comprehension of notebooks. To check this hypothesis, we performed a human evaluation of our algorithm: we re-split a small sample of the dataset and asked potential users to choose which split they found to be more appropriate.  

Firstly, we showed the participants the entire code of the notebook in the form of a script (\ie a regular Python file). This step was necessary for reducing the cognitive load bias of the first-time code viewing. Additionally, we asked \textit{``How well do you understand the script?''} to control the quality of answers. All evaluations with low scores on this question were excluded from the results. After the participants had acknowledged the code, we showed them two notebooks --- the original and the re-split. After the participants went through both notebooks, we asked them to choose which form of the code representation was the most appropriate --- the script, the first notebook, or the second one. Finally, we asked them if they wanted to split the notebook in an entirely different way. 

We employed a blind procedure for this experiment. We randomly sampled five pairs of notebooks (re-split and original) from our dataset. We used two criteria for sampling: (1)~the number of cells in a notebook should be between 5 and 15 in order to avoid edge cases of extremely short or long notebooks, and (2) the number of cells should change after re-splitting. We randomized the order of notebooks in each pair, the information about the true source of the notebooks was not accessible to the authors of the paper until the final analysis of the data. We conducted this research using a Google Form and a remotely hosted Jupyter server for viewing demonstration. 

In total, 23 developers participated in our study. 
The mean experience in software engineering for our participants was 4.76 years (SD = 2.84). 43\% of them use notebooks less than one time in a week, 39\% --- two or three times a week, and 17\% use them every day.

Participants needed to choose between notebooks in 5 tasks, resulting in 115 choices in our sample. We discarded 3 answers due to low scores of understanding the given script. As a result, we got 70 cases (62.5\%) where the participants chose the original notebooks, and 33 cases (29.5\%) when they chose a re-split variant. Also, the participants preferred the script to notebooks 9 times (8\%). 

Statistical analysis showed that the participants' choice is not independent of the task (Pearson \(\chi^2,~p < 0.05\)), so let us take a closer look at the notebooks and the preference ratio for each of them. The details of the participants' choices are presented in \Cref{table:choices}.

\begin{table}[h]
\centering
\begin{tabular}{cccc}
\toprule
\textbf{Task No.} & \textbf{Original} & \textbf{Re-split} & \textbf{Script} \\ \midrule
1    & 15 (68.2\%)       & 6 (27.3\%)           & 1 (4.5\%)      \\
2    & 15 (68.2\%)       & 6 (27.3\%)           & 1 (4.5\%)      \\
3    & 9 (39.1\%)        & 8 (34.8\%)           & 6 (26.1\%)      \\
4    & 11 (50\%)       & 11 (50\%)              & 0 (0\%)      \\
5    & 20 (87\%)       & 2 (8.7\%)           & 1 (4.3\%)      \\
\bottomrule
\end{tabular}
\vspace{0.3cm}
\caption{The number of choices between the original, the re-split, and the script views of the notebooks for each task.}
\label{table:choices}
\vspace{-0.4cm}
\end{table}

\subsubsection{Task 1} The original representation was chosen in 68.2\% of cases. The reason might be that in this notebook, our algorithm only merged cells, and it merged a lot of statements that have output, like \texttt{df.head()} or \texttt{plt.show()}.

\subsubsection{Task 2} Similar to the previous task, the original representation was chosen in 68.2\% of cases. However, in this notebook, our algorithm only split cells. We suppose that in this case, \toolname changed the original logical structure that was conveyed with the chucks of code separated by new lines, and made it less comprehensible.

\subsubsection{Task 3} This very short notebook received 26.1\% of the votes for the form of the script. Once again, our algorithm merged cells with output functions. However, in the form of a script, these statements would not have any effect either way, and still some participants wanted to read the code that way. 

\subsubsection{Task 4} This is our most successful notebook --- the re-split version was chosen in half of the cases. Here, our idea of merging worked successfully --- we merged the first three cells, in which the author introduced objects that were used later. Catching such cases was one of our main goals. 

\subsubsection{Task 5} Our least successful notebook --- the re-split version was chosen only in 8.7\% of cases. The low score could be caused by the merge action: according to our idea, we merged two mostly intra-linked cells --- one with creating the dataset object, and another one with the model object. While our algorithm did this correctly, from the data science perspective, these are two separate actions. 

Based on the results of the evaluation, we can conclude that our algorithm shows potential and works as expected in some cases, but needs to be improved with the following features:
\begin{itemize}
    \item We need to analyse statements that produce an output --- they frequently dictate the structure of the notebook and must be accounted for.
    \item We need to be more careful with the original formatting or even use it as an additional heuristic. Specifically, some users like to format the code into chunks by using newlines --- and this can happen within a single cell. 
    \item We need to pick candidates for transformation better. Currently, we do it based solely on the size of the cell and do not try to score the content of the cell.

\end{itemize}

\section{Conclusion}\label{sec:conclusion}

The task of re-splitting notebooks is an important practical task --- with reported low levels of reproducibility~\cite{pimentel2019large}, it is crucial to develop tools that could increase the comprehensibility of the medium. 

We suppose that a notebook is a distinct code entity that lies between the first sketch solution for the problem and a production-ready one. It requires its own style guide, because such things as multiple outputs or cell size are not regulated in the existing ones. 
Pimentel et al.~\cite{pimentel2021understanding} already proposed a simple linter for Jupyter notebooks, and we believe that the logic of splitting cells could be a good part for such linter.   

Also, as we stated in Section~\ref{sec:background}, we propose to perceive notebook cells as proto-functions. We believe that studying cell splits could shed some light on how people think when solving programming problems.  
Fan et al.~\cite{fan2010effects} showed that people use spaces between chunks of code for navigation in the codebase. We suggest that cell division plays the same role, but it is less controlled by style guides, so it could provide more insight into the underlying logical structure of the code.  

In this paper, we presented \toolname, an algorithm for re-splitting Jupyter notebooks without any input from the user. Our evaluation showed that in 29.5\% of cases, people preferred the restructured notebooks. Although this is not a prevailing result, we can say that our heuristic caught some of the desirable structure of notebook. We provided the detailed case studies of specific tasks in our survey and came up with a list of improvements that can make the approach even better. 

Apart from these improvements, our main target is implementing \toolname as a plugin for Jupyter notebooks. One of the main advantages of our work is that our algorithm can work during the process of developing a notebook --- deploying it as an interactive tool for data scientists is the first priority for us, since this will allow us to gather a lot of crucial feedback.
We are considering developing the plugin for the IntelliJ Platform,\footnote{The IntelliJ Platform SDK Documentation:~\url{https://plugins.jetbrains.com/docs/intellij/welcome.html}} since this will allow us to use more complex representation of code and give access to such tools as the complete data and control flow graphs, which can improve both parts of \toolname: merging and splitting. A complete data flow graph will help us to acquire more precise information about the way the notebook cells are linked, and the control flow graph would improve the splitting of cells into more logical and more efficient chunks.  

Lastly, \toolname uses a lot of hyper-parameters --- the minimum and maximum length for cells, the minimum chain length for split, etc. An important part of the future work for us is to experiment with the hyper-parameters and find the most suitable ones.

The source code of \toolname and supplementary materials are available: \url{https://github.com/JetBrains-Research/ReSplit}.

\bibliographystyle{IEEEtran}
\bibliography{IEEEabrv,paper}

\end{document}